\documentclass[english,twoside,a4paper,10pt]{article}

\usepackage[latin1]{inputenc}
\usepackage[T1]{fontenc}
\usepackage{amsmath}
\usepackage{amsfonts}
\usepackage{graphicx}
\usepackage{subfigure}
\usepackage{a4wide}
\usepackage{amssymb}
\usepackage{fancyhdr}
\usepackage{mathrsfs}
\usepackage[toc,page]{appendix}

\begin{document}

\title{\bf Spherically symmetric static vacuum solutions in Mimetic gravity}
\author{ 
Ratbay Myrzakulov\footnote{Email: rmyrzakulov@gmail.com},\,\,\,
Lorenzo Sebastiani\footnote{E-mail address: l.sebastiani@science.unitn.it
}
\\
\\
\begin{small}
Department of General \& Theoretical Physics and Eurasian Center for
\end{small}\\
\begin{small} 
Theoretical Physics, Eurasian National University, Astana 010008, Kazakhstan
\end{small}\\
}

\date{}

\maketitle

\begin{abstract}
In this paper we analyze spherically symmetric static vacuum solutions with various topologies in mimetic gravity.
When the Einstein's tensor is different from zero, a new class of solutions different from the Schwarzschild one emerges from the theory. We analyze the feature of the new solutions and we study the planar motion for the spherical case.
\end{abstract}

\maketitle

\def\thesection{\Roman{section}}
\def\theequation{\Roman{section}.\arabic{equation}}

\section{Introduction}
Last year, new approach to the dark matter issue has been proposed by Chamseddin and Mukhanov 
in such a way that 
the conformal degree of
freedom in Einstein's theory is isolated in a covariant way: the physical metric is expressed 
in terms of an auxiliary metric and a scalar field, rendering the theory invariant respect to the conformal transformations of the auxiliary metric. When the Einstein's tensor is equal to the matter stress-energy tensor, we recover all the results of General Relativity, otherwise a wide class of new solutions is admitted. In  Ref.~\cite{m1} it has been demonstrated that on flat Friedmann-Robertson-Walker metric the model can reproduce the standard matter without including it in the stress energy tensor and leads to a natural dark matter description. 
In Ref.~\cite{m2} the results have been extended to other cosmological scenarios, namely bounce cosmology, inflation, dark energy epoch. 
Cosmological 
perturbations theory for mimetic gravity was studied in Ref.~\cite{Odm00}.
In Ref.~\cite{OdMim} mimetic gravity has been analyzed in the framework of $F(R)$-modified gravity 
and in Refs.~\cite{DavoodMim1, DavoodMim2} the energy conditions for this kind of models have been investigated. 
In Ref.~\cite{Smimetic} disformal transformations in mimetic gravity have been analyzed.
Other works about the so called ``mimetic gravity'' can be found in Refs.~\cite{othermimetic1}--\cite{othermimetic4}.

In this work, we would like to analyze static spherically symmetric solutions different from the Schwarzschild one in mimetic gravity. We will consider general topological solutions and focus our attention on the vacuum case, when the Einstein's tensor is different from zero. The existence of this kind of solutions is quite interesting, since they are used to describe the exterior part of compact objects like stars, planets...therefore, it looks that in mimetic gravity new phenomenologies and graviational effects may appear.

The paper is organized as follows. In Section {\bf 2}, we will revisit the formalism of mimetic gravity and its application to dark matter. In Section {\bf 3}, we will study (pseudo)-spherically symmetric static vacuum solutions with various topologies. 
Beyond the Schwarzschild solution, the model admits a new class of solutions with Einstein's tensor different to zero. The most interesting case from cosmological point of view is the spherical one. Thus, Section {\bf 4} is devoted to the analysis of the planar motion and the circular orbits for the spherical solution. In Section {\bf 5}, we will give some comments about the cosmological constant case: if we introduce a cosmological constant in the theory, we can recover the Schwarzschild de Sitter space-time, but also in this case other solutions may emerge. Conclusions and final remarks are given in Section {\bf 6}.

\section{Mimetic gravity}

In the seminal paper of Chamseddin and Mukhanov~\cite{m1}, the theory of Einstein has been modified by redefining the metric in terms of an auxiliary metric and a scalar field, namely
\begin{equation}
g_{\mu\nu}=-\tilde g_{\mu\nu}\tilde g^{\alpha\beta}\partial_\alpha\phi\partial_\beta\phi\,,
\label{metricphi}
\end{equation}
where $\tilde g_{\mu\nu}$ is the auxiliary metric and $\phi$ is a scalar field which appears throught its first derivative. Thus, the conformal degree of freedom is explicitly isolated in a covariant way, since the physical metric (and therefore the resulting action of the theory) is invariant under conformal transformations of $\tilde g_{\mu\nu}$ as $\tilde g'_{\mu\nu}=\Omega(t, {\bf x})^2g_{\mu\nu}$, $\Omega(t, {\bf x})$ being a generic function of the coordinates.
We immediatly have
\begin{equation}
g^{\mu\nu}\partial_\mu\phi\partial_\nu\phi=-1\,.\label{phiconst}
\end{equation} 
The corresponding action of General Relativity can be rewritten as
\begin{equation}
I=\int_\mathcal{M}\left(\frac{R(g_{\mu\nu})}{16\pi G_N}+\mathcal L_m\right)\sqrt{-g\left(\tilde g_{\mu\nu},\phi\right)}dx^4\,,
\end{equation}
where $\mathcal M$ is the space-time manifold, $R\equiv R(g_{\mu\nu})$ is the Ricci tensor, $\mathcal L_m$ is the matter Lagrangian and the metric $g_{\mu\nu}$ is a function of the auxiliary metric  $\tilde g_{\mu\nu}$ and the field $\phi$ as in (\ref{metricphi}). The variation with respect to $\tilde g_{\mu\nu}$ and $\phi$ leads to
\begin{equation}
G_{\mu\nu}=8\pi G_N\left(T_{\mu\nu}+\tilde T_{\mu\nu}\right)\,,\quad\nabla^\mu \tilde T_{\mu\nu}=0\,,\label{fieldeq}
\end{equation}
where
\begin{equation}
\tilde T_{\mu\nu}=-\frac{1}{8\pi G_N}\left(G-8\pi G_N T\right)\partial_\mu\phi\partial_\nu\phi\,.\label{tildeT}
\end{equation}
In the above expressions, $G_{\mu\nu}=R_{\mu\nu}-g_{\mu\nu}R/2$ is the Einstein's tensor, $R_{\mu\nu}$ being the Riemann tensor, and $G\equiv -R$ and $T$ are the traces of the Einstein's tensor and the tensor $\tilde T_{\mu\nu}$ in (\ref{tildeT}), respectively. Moreover, $T_{\mu\nu}$ is the stress energy tensor of standard matter whose covariant derivative reads $\nabla^\mu T_{\mu\nu}=0$. Note that the trace of the first equation in (\ref{fieldeq}) leads to
\begin{equation}
(G-8\pi G_N T)(1+g^{\mu\nu}\partial_\mu\phi\partial_\nu\phi)=0\,,
\end{equation}
which is automatically satisfied if we take into account (\ref{phiconst}), even when $(G-8\pi G_N T)\neq 0$. Finally, the conservation law of the tensor $\tilde T_{\mu\nu}$ with  (\ref{phiconst}) results to be
\begin{equation}
\nabla^\kappa\left((G-8\pi G_N T)\partial_\kappa\phi\right)\equiv\frac{1}{\sqrt{-g}}
\partial_\kappa\left(\sqrt{-g}(G-8\pi G_N T)g^{\kappa\sigma}\partial_\sigma\phi\right)=0\,.\label{phieq}
\end{equation}
Given the physical metric $g_{\mu\nu}$, we can derive the form of $\phi$ from equation (\ref{phiconst}) and from the field equations (\ref{fieldeq}). Therefore, it is natural to recover all the results of the theory of Einstein when $G_{\mu\nu}=8\pi G_N T_{\mu\nu}$ and $\tilde T_{\mu\nu}=0$, but, since equation (\ref{phiconst}) alone does not fix in an unique way the field $a\,\,priori$, in deriving the field equations we also get some additional part from the freedom degree associated to the field.
In Ref.~\cite{m1} it has been demonstrated that such a freedom degree is dynamic even in the absence of matter ($T_{\mu\nu}=0$) and it is a good candidate for the dark matter. The stress energy-tensor of a perfect fluid reads
\begin{equation}
T^{\mu\nu}=(\rho+p)u^\mu u^\nu+p g^{\mu\nu}\,,\quad u_\mu u^\mu=-1\,.\label{stresspf}
\end{equation}
Thus, if we compare expression (\ref{tildeT}) with the stress-energy tensor of matter
with $p=0$, after the identification $\rho=-(G-8\pi G_N T)/8\pi G_N$ and $u_\mu=\partial_\mu\phi$, we obtain a matter-like contribute.

In Friedmann universe, whose homogeneous and isotropic metric assumes the classical form
\begin{equation}
ds^2=-dt^2+a(t)^2d{\bf x^2}\,,\quad \phi=t\,,
\end{equation}
where $a(t)$ is the scale factor depending on cosmological time and we have used conition (\ref{phiconst}), Eq.~(\ref{phieq}) reads
\begin{equation}
\partial_t\left(a^3(G-8\pi G_N T)\right)=0\,,\quad (G- 8\pi G_N T)=\frac{c_0}{a^3}\,,\label{mlaw}
\end{equation}
where $c_0$ is an integration constant: the dark matter desroption follows from the setting of a suitable value of $c_0$.

In the next section, we will consider stationary speherically symmetric vacuum solutions. In this case, the symmetry of the system imposes that the only non-zero component of the vector $\partial_\mu\phi$ is the (1,1)-component. It is clear that in order to satisfy (\ref{phiconst}) and in order to get a time-like vector, such a component must be immaginary (but the physical metric is real, thanks to the fact that also the auxiliary metric is immaginary), and the correspondence with the dark matter is only formal, since the four-velocity vector $u_\mu$ in (\ref{stresspf}) cannot be physical.

In Ref.~\cite{m2} it has also been demostrated that by introducing a vector field we obtain the dark energy able to support the current acceleration of our universe, and by using some suitable potentials we can in principle reproduce many other cosmological scenarios (bounce solutions, inflation, ...). 

In the next section, we will investigate the spherically symmetric static vacuum solutions of the model, and we will see which kind of results we can obtain when the Ricci tensor is different to zero.

\section{Spherically symmetric static vacuum solutions}

We will look for topological static, (pseudo-)spherically symmetric solutions (SSS) of the type
\begin{equation}
ds^2=-e^{2\alpha(r)}B(r)dt^2+\frac{d r^2}{B(r)}+r^2\,\left(\frac{d\rho^2}{1-k\rho^2}+\rho^2 d\phi^2\right)\label{SSS}\,,
\end{equation}
where $\alpha(r)\,, B(r)$ are general functions of the radial coordinaye $r$, and the manifold is a sphere when $k=1$, a torus when $k=0$ or a compact hyperbolic manifold when $k=-1$. Thus, the Ricci scalar is derived as
\begin{eqnarray}
R  &=&
-\frac{1}{r^2}\left[3r^2\,\left({\frac{d B(r)}{dr}}\right)\left({\frac{d\alpha(r)}{dr}}\right)
+2r^2\,B\left(r\right)\left({\frac{d\alpha(r)}{dr}}
\right)^{2}+r^2\left({\frac{d^{2}B(r)}{d{r}^{2}}}\right)
\right.\nonumber\\ \nonumber\\
&&+2r^2\,B\left(r\right)\left({\frac{d^{2}\alpha(r)}{d{r}^{2}}}\right)\left.+4r\,\left({\frac{d B(r)}{dr}}\right)
+4 r B(r)\,\left({\frac{d\alpha(r)}{dr}}\right)+2B(r)
-2k\right]\,.\label{SSSRicci}
\end{eqnarray}
First of all, we observe that
the topological Schwarzschild de Sitter solution 
\begin{equation}
B(r)=k-\frac{M}{r}-\frac{\Lambda r^2}{3}\,,\quad\alpha(r)=\text{const}\,,\label{SdSsol}
\end{equation}
where $M$ is an integration (mass) constant and $\Lambda$ is a cosmological constant, is a solution of the model in (\ref{fieldeq}) for $T_{\mu\nu}=-\Lambda g_{\mu\nu}/(8\pi G_N)$ with $G_{\mu\nu}=8\pi G_N T_{\mu\nu}$. In such a case, $\tilde T_{\mu\nu}=0$ and the results of General Relativity are recovered. 

Now, it may be interesting to see if other types of SSS solutions appear when
$G_{\mu\nu}\neq 8\pi G_N T_{\mu\nu}$. 
We will consider the vacuum case ($T_{\mu\nu}=0$ but $G_{\mu\nu}\neq 0$), for which equations~(\ref{phiconst})--(\ref{tildeT}) read
\begin{equation}
R_{\mu\nu}-\frac{g_{\mu\nu}R}{2}=R\partial_\mu\phi\partial_\nu\phi\,,\quad
\partial_\kappa\left(\sqrt{-g}R g^{\kappa\sigma}\partial_\sigma\phi\right)=0\,,\label{setSSS}
\end{equation}
where we have used (\ref{phieq}). 
From (\ref{phiconst}), by assuming the SSS geometry, it follows
\begin{equation}
\frac{d\phi}{d r}=\sqrt{-\frac{1}{B(r)}}\,,\label{fieldSSS}
\end{equation}
and $\phi$ is an imaginary field. Therefore, the field equations of the model read
\begin{equation}
\frac{d B(r)}{d r}r+B(r)-k=0\,,\label{uno}
\end{equation}
\begin{eqnarray}
&&\left[3r^2\,\left({\frac{d B(r)}{dr}}\right)\left({\frac{d\alpha(r)}{dr}}\right)
+2r^2\,B\left(r\right)\left({\frac{d\alpha(r)}{dr}}
\right)^{2}+r^2\left({\frac{d^{2}B(r)}{d{r}^{2}}}\right)
\right.\nonumber\\ \nonumber\\
&&+2r^2\,B\left(r\right)\left({\frac{d^{2}\alpha(r)}{d{r}^{2}}}\right)\left.+3r\,\left({\frac{d B(r)}{dr}}\right)
+2 r B(r)\,\left({\frac{d\alpha(r)}{dr}}\right)+B(r)
-k\right]=0\,,\label{due}
\end{eqnarray}
\begin{eqnarray}
\hspace{-1.5cm}2\frac{d B(r)}{d r}-2B(r)\frac{d\alpha(r)}{d r}+r\frac{d^2 B(r)}{d r^2}+3r\frac{d B(r)}{d r}\frac{d\alpha(r)}{d r}
+2r B(r)\frac{d^2\alpha(r)}{d r^2}+2r B(r)\left(\frac{d\alpha(r)}{d r}\right)^2=0\,,
\end{eqnarray}
where the last equation corresponds to $G_{22}=G_{33}=0$ and it is automatically satisfied if the first two equations hold true. Equation (\ref{uno}) leads to
\begin{equation}
B(r)=k-\frac{M}{r}\,,\label{B}
\end{equation}
$M$ being a mass term like in Schwarzschild metric. By plugging this expression in Eq.~(\ref{due}), one has
\begin{eqnarray}
\hspace{-1.5cm}\left[3r^2\,\left({\frac{d B(r)}{dr}}\right)\left({\frac{d\alpha(r)}{dr}}\right)
+2r^2\,B\left(r\right)\left({\frac{d\alpha(r)}{dr}}
\right)^{2}
+2r^2\,B\left(r\right)\left({\frac{d^{2}\alpha(r)}{d{r}^{2}}}\right)+2r B(r)\left(\frac{d\alpha(r)}{d r}\right)
\right]=0\,,\label{aaaa}
\end{eqnarray}
whose solution can be written as
\begin{equation}
\exp[2\alpha(r)]=\pm\frac{A_0}{\left(1-\frac{k M}{r}\right)}\left[
\left(\sqrt{1-\frac{k M}{r}}\right)
\log\left[\sqrt{\frac{r}{r_0}}\left(1+\sqrt{1-\frac{k M}{r}}\right)\right]-1\right]^2\,,\quad k=-1\,,1\,,\label{k1}
\end{equation}
\begin{equation}
\exp[2\alpha(r)]=A_0\left[2\left(\frac{r}{r_0}\right)^{3/2}+3\right]^2\,,\quad k=0\,,\label{k0}
\end{equation}
where $0<A_0\,,r_0$ are integration constants (usually one deals with the gauge $A_0=1$). Note that also the second equation in (\ref{setSSS}) is satisfied.\\
\\
Let us analyze the results. 
The solution for flat topology ($k=0$) assumes the following form,
\begin{equation}
ds^2=-A_0\left[2\left(\frac{r}{r_0}\right)^{3/2}+3\right]^2\left(\frac{\tilde M}{r}\right)dt^2+
\frac{dr^2}{\left(\frac{\tilde M}{r}\right)}+r^2\left(d\rho^2+\rho^2 d\phi^2\right)\,,\quad 0<\tilde M\,,
\end{equation}
where we have introduced $0<\tilde M=-M$ to preserve the metric signature. In this case the Ricci scalar is different to zero and reads
\begin{equation}
R=-\frac{6\tilde M}{2r^3+3\left(\frac{r}{r_0}\right)^{3/2}r_0^3}\,.
\end{equation}
Thus, we have a naked singularity at $r=0$, like for the corresponding case $k=0$ in Schwarzshild solution.

The solution for the spherical case ($k=1$) is the more interesting from cosmological point of view and is given by
\begin{equation}
ds^2=-A_0\left[
\left(\sqrt{1-\frac{M}{r}}\right)
\log\left[\sqrt{\frac{r}{r_0}}\left(1+\sqrt{1-\frac{M}{r}}\right)\right]-1\right]^2 dt^2
+\frac{dr^2}{\left(1-\frac{M}{r}\right)}+r^2\left(d\theta^2+\sin^2\theta d\phi^2\right)\,,
\label{k1}
\end{equation}
where we have introduced the polar coordinates $\theta\,,\phi$ in the angular part. This result is in agreement with Ref.~\cite{Smimetic}. The Ricci scalar results to be
\begin{equation}
R=-\frac{1}{r^2\left[
\left(\sqrt{1-\frac{M}{r}}\right)
\log\left[\sqrt{\frac{r}{r_0}}\left(1+\sqrt{1-\frac{M}{r}}\right)\right]-1\right]}\,.
\end{equation}
If $M<0$, the metric is regular everywhere.

In the case of $0<M$, the metric coefficients $g_{00}(r)\,, g_{11}(r)$ are positive for $M<r$, while for $r<M$ we have $g_{11}(r)<0$ and $g_{00}(r)$ aquires an imaginary part. A special case is represented by the choice $M=r_0$, for which when $r<M$ we get
\begin{equation}
g_{00}(r)=-A_0\left(\frac{M}{r}\right)^2\,,
\end{equation}
and $g_{00}(r)$ is real and negative.

When $r=M$, $g_{11}(r)$ diverges but $g_{00}(r)=-A_0$ and the Ricci scalar is regular. It follows that $r=M$ does not represent an ``horizon'' (we also cannot associate to it any thermodynamical quantity like the temperature $T=(\exp[\alpha(r)]/2) d B(r)/dr$ which diverges at $r=M$). Since the solution becomes imaginary for $r<M$, we must consider the branch $M<r$ only.
We avoid to discuss the special case $r_0=M$, for which when we cross the point $r=M$ the signature of $g_{11}(r)$ changes in $(--++)$.

Finally, the topological case $k=-1$ corresponds to
\begin{equation}
ds^2=-A_0 \left[
\left(\sqrt{1+\frac{M}{r}}\right)
\log\left[\sqrt{\frac{r}{r_0}}\left(1+\sqrt{1+\frac{M}{r}}\right)\right]-1\right]^2 dt^2
+\frac{dr^2}{\left(-1-\frac{M}{r}\right)}+r^2\,\left(\frac{d\rho^2}{1+\rho^2}+\rho^2 d\phi^2\right)\,,\label{k-1}
\end{equation}
and the Ricci scalar reads
\begin{equation}
R=-\frac{1}{r^2\left[
\left(\sqrt{1+\frac{M}{r}}\right)
\log\left[\sqrt{\frac{r}{r_0}}\left(1+\sqrt{1+\frac{M}{r}}\right)\right]-1\right]}\,.
\end{equation}
Now, if $0<M$, the metric coefficent $g_{11}(r)$ is negative and the solution is unphysical.
On the other hand, by choosing $M<0$, we have that $0<g_{11}(r)$ if $r<-M$, but $g_{00}(r)$ becomes imaginary (except for the special choice $r_0=-M$). 

In the next section, we will analyze the planar orbits for the spherical case $k=1$ assuming $M<r$ (if $M<0$, the radial coordinate covers all the space).

\section{Planar motion for spherical solution}

Let us analyze the motion of a free particle in the gravitational field of solution (\ref{k1}) with $k=1$, namely the case with spherical topology, which is the most interesting from cosmological point of view.
The motion of a free particle is governed by the geodesic equation,
\begin{equation}
\frac{d^2 x^\mu}{d\tau^2}+\Gamma^{\mu}_{\alpha\beta}\frac{d x^\alpha}{d\tau}\frac{d x^\beta}{d\tau}=0\,,\label{geo}
\end{equation}
where $\tau$ denotes the proper time and $\Gamma^\mu_{\alpha\beta}$ are the Christoffel symbols,
\begin{equation}
\Gamma^\mu_{\alpha\beta}=\frac{g^{\mu\sigma}}{2}\left(\partial_\alpha g_{\sigma\beta}+\partial_\beta g_{\sigma\alpha}-\partial_\sigma g_{\alpha\beta}\right)\,.
\end{equation}
Given the metric (\ref{k1}), the non-zero Christoffel symbols read
\begin{equation}
\Gamma^0_{01}=\Gamma^0_{10}=\frac{r-M+Y(X+1)M+X r}{2(1+X)X r^2(XY-1)}\,,\nonumber
\end{equation}
\begin{equation}
\Gamma^{1}_{00}=-\frac{A_0(M-r)(XY-1)}{2r^2(r-M+Xr)}\left(r-M+Xr+MY(1+X)\right)\,,\nonumber
\end{equation}
\begin{equation}
\Gamma^1_{11}=\frac{M}{2(M-r)r}\,,\quad
\Gamma^1_{22}=(M-r)\,,\quad
\Gamma^1_{33}=(M-r)\sin^2\theta\,,\nonumber
\end{equation}
\begin{equation}
\Gamma^2_{12}=\Gamma^2_{21}=\frac{1}{r}\,,\quad\Gamma^2_{33}=-\sin\theta\cos\theta\,,
\nonumber
\end{equation}
\begin{equation}
\Gamma^3_{13}=\Gamma^3_{31}=\frac{1}{r}\,,\quad
\Gamma^3_{23}=\Gamma^3_{32}=\frac{\cos\theta}{\sin\theta}\,,
\end{equation}
where
\begin{equation}
X=\sqrt{1-\frac{M}{r}}\,,\quad
Y=\log\left[\sqrt{\frac{r}{r_0}}\left(1+\sqrt{1-\frac{M}{r}}\right)\right]\,.\label{XY}
\end{equation}
Thus, equation (\ref{geo}) corresponds to
\begin{equation}
\frac{d}{d\tau}\left(\frac{dt}{d\tau}\right)+2\Gamma^{0}_{01}\left(\frac{dt}{d\tau}\right)
\left(\frac{dr}{d\tau}\right)=0\,,\nonumber
\end{equation}
\begin{equation}
\frac{d}{d\tau}\left(\frac{dr}{d\tau}\right)+\Gamma^{1}_{00}\left(\frac{dt}{d\tau}\right)
\left(\frac{dt}{d\tau}\right)
+\Gamma^{1}_{11}\left(\frac{dr}{d\tau}\right)
\left(\frac{dr}{d\tau}\right)
+\Gamma^{1}_{22}\left(\frac{d\theta}{d\tau}\right)
\left(\frac{d\theta}{d\tau}\right)
+\Gamma^{1}_{33}\left(\frac{d\phi}{d\tau}\right)
\left(\frac{d\phi}{d\tau}\right)=0\,,\nonumber
\end{equation}
\begin{equation}
\frac{d}{d\tau}\left(\frac{d\theta}{d\tau}\right)+2\Gamma^{2}_{12}\left(\frac{dr}{d\tau}\right)
\left(\frac{d\theta}{d\tau}\right)
+\Gamma^{2}_{33}\left(\frac{d\phi}{d\tau}\right)
\left(\frac{d\phi}{d\tau}\right)=0\,,\nonumber
\end{equation}
\begin{equation}
\frac{d}{d\tau}\left(\frac{d\phi}{d\tau}\right)+2\Gamma^{3}_{13}\left(\frac{dr}{d\tau}\right)
\left(\frac{d\phi}{d\tau}\right)
+2\Gamma^{3}_{23}\left(\frac{d\theta}{d\tau}\right)
\left(\frac{d\phi}{d\tau}\right)=0\,.\label{sist}
\end{equation}
From (\ref{k1}) one has
\begin{eqnarray}
\left(\frac{ds}{d\tau}\right)^2&=&-A_0\left[
\left(\sqrt{1-\frac{M}{r}}\right)
\log\left[\sqrt{\frac{r}{r_0}}\left(1+\sqrt{1-\frac{M}{r}}\right)\right]-1\right]^2 \left(\frac{dt}{d\tau}\right)^2
+\frac{1}{\left(1-\frac{M}{r}\right)}\left(\frac{dr}{d\tau}\right)^2
\nonumber\\&&+r^2\left(\left(\frac{d\theta}{d\tau}\right)^2+\sin^2\theta \left(\frac{d\phi}{d\tau}\right)^2\right)\,.
\label{kk1}
\end{eqnarray}
Since in the presence of a spherical symmetry the orbits remain on a plane, we can fix $\theta=\pi/2$ and $(d\theta/d\tau)=0$, such that we immediatly get from the first, third and fourth equations in the system above (\ref{sist}),
\begin{equation}
\left(\frac{dt}{d\tau}\right)=
\frac{t_0}{\left[
\left(\sqrt{1-\frac{M}{r}}\right)
\log\left[\sqrt{\frac{r}{r_0}}\left(1+\sqrt{1-\frac{M}{r}}\right)\right]-1\right]^2}\,,\label{t}
\end{equation}
\begin{equation}
\frac{d}{d\tau}\left(\frac{d\theta}{d\tau}\right)= 0\,,\label{theta}
\end{equation}
\begin{equation}
\left(\frac{d\phi}{d\tau}\right)=\frac{l_0}{r^2}\,,\label{phi}
\end{equation}
where $t_0$ and $l_0$ are constants. The last equation expresses the conservation of the angular momentum.\\ 
\\
For time-like orbits, since $ds^2=-d\tau^2$, we get from (\ref{kk1}) with (\ref{t})--(\ref{phi}),
\begin{equation}
-1=-\frac{A_0 t_0}{\left[\left(\sqrt{1-\frac{M}{r}}\right)\log\left[\sqrt{\frac{r}{r_0}}\left(1+\sqrt{1-\frac{M}{r}}\right)\right]-1\right]^2}+\frac{1}{\left(1-\frac{M}{r}\right)}\left(\frac{dr}{d\tau}\right)^2+\frac{l_0^2}{r^2}\,.\label{timelike}
\end{equation}
It follows,
\begin{equation}
\left(\frac{dr}{d\tau}\right)^2=
\left[-1+\frac{A_0 t_0}{(XY-1)^2}-\frac{l_0^2}{r^2}
\right]\left(1-\frac{M}{r}\right)\,,
\end{equation}
where we have used (\ref{XY}).
To obtain the equation of the orbit we must divide the both sides of this expression by $(d\phi/d\tau)$ in (\ref{phi}), namely,
\begin{equation}
\left(\frac{dr}{d\phi}\right)^2=\left[-1+\frac{A_0 t_0}{(XY-1)^2}-\frac{l_0^2}{r^2}
\right]\left(\frac{r^4}{l_0^2}-\frac{Mr^3}{l_0^2}\right)\,,
\end{equation}
or, by introducing $u=1/r$,
\begin{equation}
\left(\frac{du}{d\phi}\right)^2=
\left[-\frac{1}{l_0^2}+\frac{M u}{l_0^2}-u^2+M u^3\right]
+\frac{A_0 t_0}{(XY-1)^2}\left(\frac{1}{l_0^2}-\frac{M u}{l_0^2}\right)\,,\quad
u=\frac{1}{r}\,,\label{geod}
\end{equation}
and
\begin{equation}
XY=\sqrt{1-Mu}\log\left[\sqrt{\frac{u_0}{u}}\left(1+\sqrt{1-Mu}\right)\right]\,,\quad u_0=\frac{1}{r_0}\,.
\end{equation}
Equation (\ref{geod}) describes the orbital motion for SSS solution (\ref{k1}) with spherical topology. 
An important application is the 
study of the circular motion: given $u$ with the gauge $A_0$ and the mass constant $M$, if $XY\neq 1$, we can fix $t_0$ or $l_0$ to obtain the condition $(du/d\phi)=0$.
For example, in the asymptotic limit
$u\rightarrow 0^+$, namely $r\rightarrow+\infty$, we have to require $|1/l_0|\ll 1$ and $(d\phi/d\tau)$ in Eq.~(\ref{phi}) becomes very large. It means that at large distances we have high spin particles, how it is possible to understand by looking for the metric (\ref{k1}):
when $r\rightarrow\infty$, $-d\tau^2\simeq-g_{tt}(r)dt^2$ where $g_{tt}(r\rightarrow+\infty)\rightarrow+\infty$, such that $dt\ll d\tau$ and the velocity of particles increases. 
We also observe that for $0<M$, when $u$ is close to $1/M$, namely near to the singularity of the metric, $(d u/d\phi)\simeq 0$ thanks to the fact that $g_{rr}(r=M)\neq0$: in this case, the particles are forced to follow circular orbits. On the other hand, when $XY=1$ (for example, if $M=1/u_0$, $XY\simeq1$ for $u\simeq 0.3 u_0$),
the derivative $(du/d\phi)$ diverges and the circular orbit cannot be realized.

One may be interested in the analysis of the stability of the circular orbits. Let us take $r=r^*$ such that
\begin{equation}
F(u^*)=0\,,\quad\frac{d F(u)}{d u}|_{u=u^*}=0\,,\quad u^*=\frac{r}{r^*}\,,
\end{equation}
where
\begin{equation}
F(u)=\left[-\frac{1}{l_0^2}+\frac{M u}{l_0^2}-u^2+M u^3\right]
+\frac{A_0 t_0}{(XY-1)^2}\left(\frac{1}{l_0^2}-\frac{M u}{l_0^2}\right)\,.
\end{equation}
From Eq.~(\ref{geod}) one has
\begin{equation}
\frac{d u}{d\phi}|_{u=u^*}=0\,,\quad \frac{d^2 u}{d \phi^2}|_{u=u^*}=\frac{1}{2}\frac{d F(u)}{d u}|_{u=u^*}=0\,.
\end{equation}
If we perturbate the second equation around $u=u^*+\Delta u\,,|\Delta u|\ll 1$, we obtain
\begin{equation}
\frac{d^2\Delta u}{d u^2}=\frac{1}{2}\frac{d^2 F(u)}{d u^2}|_{u=u^*}\Delta u=0\,,
\end{equation}
whose solution is given by
\begin{equation}
\Delta u=\delta\exp\pm\left[\frac{1}{\sqrt{2}}\sqrt{\left(\frac{d^2 F(u)}{d u^2}|_{u=u^*}\right)}\phi
\right]\,,
\end{equation}
$\delta$ being a constant. It means that, given $u=u^*$, if $0<d^2F(u^*)/d u^2$ the circular orbit is unstable, otherwise is stable. For example, in the limit $u^*\rightarrow 0$, one gets $d^2F(u^*)/d u^2\simeq -2$ and the orbit is stable for large values of $r$.\\
\\
Finally, let us consider the null geodesics with $ds^2=0$. Instead of the proper time, we use $\sigma=x^0$ and (\ref{t})--(\ref{phi}) are still valid. Then, Eq.~(\ref{timelike}) becomes
\begin{equation}
0=-\frac{A_0 t_0}{\left[\left(\sqrt{1-\frac{M}{r}}\right)\log\left[\sqrt{\frac{r}{r_0}}\left(1+\sqrt{1-\frac{M}{r}}\right)\right]-1\right]^2}+\frac{1}{\left(1-\frac{M}{r}\right)}\left(\frac{dr}{d\sigma}\right)^2+\frac{l_0^2}{r^2}\,.\label{null}
\end{equation}
By following the same approach of the time-like case and by multiplying this expression by $(d\phi/d\sigma=l_0/r^2)$, we get in terms of
$u=1/r$,
\begin{equation}
\left(\frac{du}{d\phi}\right)^2=
\left(-u^2+M u^3\right)
+\frac{A_0 t_0}{(XY-1)^2}\left(\frac{1}{l_0^2}-\frac{M u}{l_0^2}\right)\,,\quad
u=\frac{1}{r}\,.
\end{equation}
This equation describes the planar trajectory of photons in the gravitational field of the solution under consideration.
We observe that the circular orbits with $(du/d\phi)=0$ are always realized for $u\simeq1/M\,,0<M$ and $u\rightarrow 0^+$.

\section{Notes on the solutions with cosmological constant}

Let us return to the SSS topological metric (\ref{SSS}) and introduce a cosmological constant $\Lambda$ in  the model, namely $T_{\mu\nu}=-\Lambda g_{\mu\nu}/(8\pi G_N)$ in (\ref{fieldeq}). If $G_{\mu\nu}=8\pi G_N T_{\mu\nu}$, we recover the Schwarzschild de Sitter space-time (\ref{SdSsol}), but when $G_{\mu\nu}\neq 8\pi G_N T_{\mu\nu}$
equations~(\ref{phiconst})--(\ref{tildeT}) lead to
\begin{equation}
R_{\mu\nu}-\frac{g_{\mu\nu}R}{2}+\Lambda g_{\mu\nu}=\left(R+8\pi G_N T\right)\partial_\mu\phi\partial_\nu\phi\,,\quad
\partial_\kappa\left(\sqrt{-g}R g^{\kappa\sigma}\partial_\sigma\phi\right)=0\,.\label{setSSS2}
\end{equation} 
Thus, by using (\ref{fieldSSS}), one finds the following solution for the metric (\ref{SSS}),
\begin{equation}
ds^2=-e^{2\alpha(r)}\left(k-\frac{M}{r}-\frac{\Lambda r^2}{3}\right)dt^2+\frac{d r^2}{\left(k-\frac{M}{r}-\frac{\Lambda r^2}{3}\right)}+r^2\,\left(\frac{d\rho^2}{1-k\rho^2}+\rho^2 d\phi^2\right)\label{SSS}\,,
\end{equation}
where $\alpha(r)$ is given by Eq.~(\ref{aaaa}) with $B(r)=k-M/r-\Lambda r^2/3$. The solution is implicit for $k=\pm 1$, but for $k=0$ one has
\begin{equation}
\alpha(r)=\frac{9 M\sqrt{r}}{6M r^{3/2}+2\Lambda r^{9/2}+c_0\sqrt{3M+r^3\Lambda}\left(27M^2+9M\Lambda r^3\right)}\,,
\end{equation}
where $c_0$ is a constant. Note that the factor $\alpha(r)$ never diverges, and the metric becomes singular on the cosmological horizon $r=(3M/\Lambda)^{1/3}$.

We can say that in the presence of a cosmological constant the metric coefficient $g_{rr}(r)$ corresponds to the Schwarzschild de Sitter one, with the appearance of an integration mass constant. Like in the vacuum case, what differs from Schwarzschild de Sitter space-time is the metric coefficient $g_{tt}(r)$, and new features can be aquired.

\section{Conclusions}

Mimetic gravity has been proposed as a theory to reproduce the dark matter phenomenology without invoking its presence in Friedmann-Robertson-Walker universe. All the results of General Relativity can be recovered, but new classes of solutions may emerge from the theory. In this paper, we have considered 
(pseudo)-spherically symmetric static vacuum solutions with various topologies. In this respect, the model admits some metrics different 
from the Schwarzschild one, and in the specific three solutions for the spherical, toroidal and hyperbolic manifolds have been derived as the only vacuum solutions of the theory when the Einstein's tensor differs from zero. The toroidal and hyperbolic manifold cases are not very interesting from the cosmological point of view, and the physical range of the hyperbolic solution is quite restricted. More interesting is the analysis of the spherical solution, where the coefficient $g_{rr}(r)$ of the metric turns out to be the Schwarzschild one with an integration mass constant $M$, but the metric coefficient $g_{tt}(r)$ reads in a different way. In particular, it is regular when the radial coordiante coincides with $M$, such that the solution does not present a real horizon at $r=M$. The solution is physical when $M<r$ (if $M<0$ it covers all the space): for example, we may describe cosmological objects whose size exceed $r=M$. We have investigated the planar motion of free particles and we have found the equation which describes the orbits at fixed angle  $\theta=\pi/2$. This equation can be used to study the circular orbits and it is possible to see that at large distances the graviational effects are very strong and the particles aquire an high spin. The orbital equation of photons has been also presented. Finally, in the last section of our work, we gave some comments about the cosmological constant case, when the introduction of a cosmological constant in the field equations of mimetic gravity may give arise to a class of solutions different from the Schwarzschild de Sitter one if the Einstein's tensor does not coincide with the stress energy tensor related to the cosmological constant.

\section*{Acknowledgments}

L.S. thanks D.Momeni for comments and useful discussions.

\end{document}